\documentclass[aps,twocolumn,showpacs,amssymb,%
floatfix,superscriptaddress]{revtex4}
\usepackage{graphicx}
\usepackage{epsf}
\usepackage{dcolumn}
\usepackage{bm}
\usepackage{hyperref}
\usepackage{latexsym}
\usepackage{color}

\begin{document}
\def\av#1{\langle#1\rangle}
\def\etal{{\it et al\/.}}
\def\pc{p_{\rm c}}
\def\l{{\lambda}}
\def\hm{h_*}
\def\xm{x_*}
\def\remark#1{{\bf *** #1 ***}}
\def\beq{\begin{equation}}
\def\eeq{\end{equation}}

\title{Wikipedias: Collaborative web-based encyclopedias as complex networks}

\author{V. Zlati\'{c}\thanks{vzlatic@irb.hr}}
\affiliation{Theoretical Physics Division, Rudjer Bo\v{s}kovi\'{c} Institute, \\
   P.O.Box 180, HR-10002 Zagreb, Croatia}
\author{M. Bo\v{z}i\v{c}evi\'c\thanks{miran@mi2.hr}}
\affiliation{Multimedia Institute, Preradovi\'{c}eva 18, HR-10000 Zagreb,
Croatia}
\affiliation{Institute for Social Research in Zagreb, Amru\v{s}eva 8, HR-10000
Zagreb, Croatia}
\author{H. \v{S}tefan\v{c}i\'{c}\thanks{shrvoje@thphys.irb.hr}}
\affiliation{Theoretical Physics Division, Rudjer Bo\v{s}kovi\'{c} Institute, \\
   P.O.Box 180, HR-10002 Zagreb, Croatia}
\author{M. Domazet \thanks{domazet@idi.hr}}
\affiliation{Institute for Social Research in Zagreb, Amru\v{s}eva 8, HR-10000
Zagreb, Croatia}

\begin{abstract}

Wikipedia is a popular web-based encyclopedia edited freely and
collaboratively by its users. In this paper we present an analysis
of Wikipedias in several languages as complex networks. The
hyperlinks pointing from one Wikipedia article to another are
treated as directed links while the articles represent the nodes of
the network. We show that many network characteristics are common to
different language versions of Wikipedia, such as their degree
distributions, growth, topology, reciprocity, clustering,
assortativity, path lengths and triad significance profiles. These
regularities, found in the ensemble of Wikipedias in different
languages and of different sizes, point to the existence of a unique
growth process. We also compare Wikipedias to other previously
studied networks.
\end{abstract}
\maketitle

\section{Introduction}

In the last few years the physics community has paid a lot of attention to the
field of complex networks. A considerable amount of research has been done on
different real world networks, complex network theory and mathematical models
\cite{AB02,DM02,N03,Strogatz}. Many real world
systems can be described as complex networks: WWW \cite{www}, internet routers
\cite{internet1,internet2,internet3}, proteins \cite{Jeong01} and
scientific collaborations \cite{GI95}, among others. Complex network theory benefitted from the study of such networks both from the
motivational aspect as well as from the new problems that arise with every newly
analyzed system.

In this paper we will present an analysis of Wikipedias in different languages
as complex networks. Wikipedia \cite{WIKI} is
a web-based encyclopedia with an unusual editorial policy that anybody can
freely edit and crosslink articles as long as one follows a simple set of
rules. Although there has been a lot of debate on the quality of Wikipedia
articles, recent findings reported in \cite{quallity} suggest that the
factographic accuracy of the English Wikipedia is not much worse than that of
the editorially compiled encyclopedias such as \textit{Encyclopaedia
Britannica}.

The important facts for this paper are: 1. that authors are encouraged to link
out of their articles, and 2. that each Wikipedia is a product of a cooperative
community. The former comes in part from the need for lexicographic links
providing context for the topic at hand, and in part from the fact that the
official Wikipedia article count, serving as the main criterion for comparing
encyclopedia sizes, includes only articles that contain an out-link. A community
arises initially from the need to follow the central Wikipedia policy of the
neutral point of view (NPOV): if there is a dispute regarding the content of an
article,
effectively all the opposing views and arguments regarding the topic should be
addressed. Although there are many occasional contributors, the bulk of the work
is done by a minority: roughly 10\% of contributors edit 80\% of the articles,
and the differing degree of authors' involvement serves as a rough criterion for
a meritocracy. Hence, there is no central structure that governs the writing of
a Wikipedia, but the process is not entirely haphazard.

We view each Wikipedia as a network with nodes corresponding to articles and
directed links corresponding to hyperlinks between them. There are over 200
Wikipedias in different languages, with different
number of nodes and links, which are continuously growing by the addition of new
nodes and creation of new links. The model of Wikipedia growth based on the
``preferential attachment" \cite{BarAlb} has been recently tested against the
empirical data \cite{wiki}. Although different Wikipedias are developed mostly independently, a number of
people have contributed in two or more different
languages, and thus participated in creating different Wikipedia networks. A
certain number of
articles have been simply translated from one language Wikipedia into
another. Also, larger Wikipedias set precedents for smaller ones on issues of
both structure and governance. There is thus a degree of interdependence between
Wikipedias in different languages. However, each language community has its
unique characteristics and idiosyncrasies, and it can be assumed that the growth
of each Wikipedia is an autonomous process, governed by the ``function affects
structure" maxim.

Namely, despite being produced by independent communities, all Wikipedias (both
in their content and in their
structure) aim to reflect the ``received knowledge" \cite{Roush},
which in general should be universal and inter-linguistic. It is expected that
community-specific
deviations of structure occur in cases where the content is less
universal than e.g. in natural science, but it is also expected that such
deviations plague each Wikipedia at some stage of its development. We
thus assume we are looking at real network realizations of different stages of
essentially
the same process of growth, implemented by different communities. By showing
which network characteristics are more
general and which more particular to individual Wikipedias and the process of
Wikipedia growth, we hope
to provide insight into generality and/or particularity of the network growth
processes.

\section{Data}
The main focus of our study is to compare networks of lexicographic articles
between different languages. However, the Wikipedia dataset is very rich, and it
is not easily reducible to a simple network in which each Wiki page is a node,
as various kinds of Wiki pages play different roles. In particular, the dataset
contains:
\begin{itemize}
\item \textit{articles,} ``normal" Wiki pages with lexicographic topics;
\item \textit{categories,} Wiki pages that serve to categorize articles;
\item \textit{images and multimedia} as pages in their own right;
\item \textit{user, help} and \textit{talk} pages;
\item \textit{redirects,} quasi-topics that simply redirect the user to another
page;
\item \textit{templates,} standardized insets of Wiki text that may add links
and categories to a page they are included in; and
\item \textit{broken links,} links to articles that have no text and do not
exist in the database, but may be created at some future time.
\end{itemize}

We studied 30 largest language Wikipedias with the data from January
7, 2005. Especially we focused on eleven largest languages as
measured by the number of undirected links. In order of size, as
measured by the number of nodes, these are: English (en), German
(de), Japanese (ja), French (fr), Swedish (sv), Polish (pl), Dutch
(nl), Spanish (es), Italian (it), Portuguese (pt) and Chinese (zh).
Based on different possible approaches to the study we analyzed six
different datasets for each language with varying policies
concerning the selection of data. We present our results for the
smallest subset we studied for each language, designed to match the
knowledge network of actual lexicographic topics most closely. It
excludes categories, images, multimedia, user, help and talk pages,
as well as broken links, and replaces redirects and templates with
direct links between articles. For a detailed explanation of the
dataset selection issues, please see our webpage
\cite{DataSelection}. An interesting measurement of the Wikipedia
dataset statistical properties is given in \cite{Voss}, and a nice
visualization of the Wikipedia data can be found in \cite{Holloway}.

\section{Results}

\subsection{Degree distribution}

\begin{figure*}
\bigskip
\includegraphics*[width=0.8\textwidth]{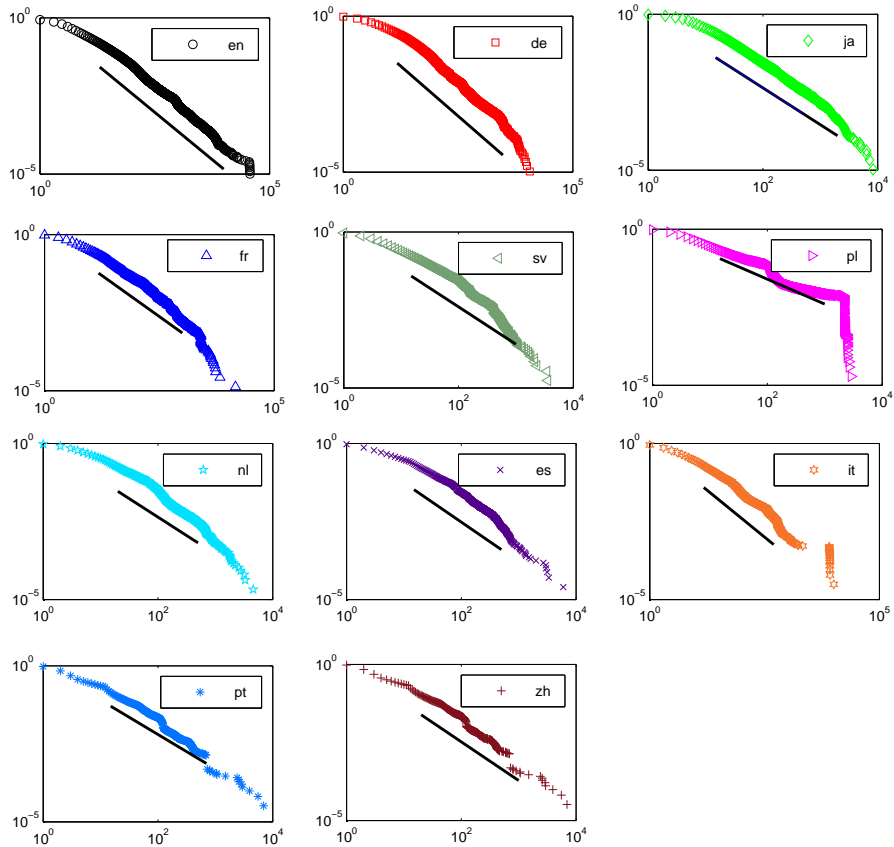}
\caption{\label{Fig:kumulativne}(Color online) This figure
represents cumulative in-degree distributions of the eleven largest
languages. In all plots in the figure the abscissa represents the node degree and 
the ordinate represents the cumulative degree distribution. The start and 
the end of the drawn best fit straight lines coincide with $k_{min}$ and 
$k_{max}$ used in fits, respectively. The power law seems applicable to all of them except
Polish. This discrepancy is related to the editorial decision of the
Polish community to heavily interlink the calendar pages using
standard templates. This community decision produced a radical
change in the structure of the network. One should also note an
unusual distribution for Italian, suggesting a similar cause.}
\end{figure*}

\begin{footnotesize}
\begin{table}[t]
\begin{center}
\begin{tabular}{|l||c|c||c|c||c|c|}
\hline
&\multicolumn{2}{c|}{in}&\multicolumn{2}{c|}{out}&\multicolumn{2}{c|}{undirected
}\\
\hline
language&$\gamma$&error&$\gamma$&error&$\gamma$&error\\
\hline\hline
en&2.21&0.04&2.65&0.15&2.37&0.04\\
\hline
de&2.28&0.05&2.60&0.15&2.45&0.05\\
\hline
ja&2.18&0.03&2.56&0.09&2.41&0.04\\
\hline
fr&2.05&0.06&2.70&0.2&2.38&0.06\\
\hline
sv&2.20&0.1&2.50&0.2&2.30&0.08\\
\hline
pl&1.80&0.1&1.80&0.2&1.85&0.09\\
\hline
nl&2.18&0.12&2.56&0.15&2.38&0.08\\
\hline
es&2.26&0.10&2.70&0.2&2.40&0.08\\
\hline
it&2.20&0.1&2.80&0.2&2.44&0.07\\
\hline
pt&2.10&0.1&2.80&0.2&2.50&0.1\\
\hline
zh&2.24&0.05&2.60&0.1&2.40&0.1\\
\hline
\hline
average&2.15&0.13&2.57&0.27&2.35&0.17\\
&2.18&0.02&2.58&0.05&2.38&0.02\\
\hline
average&2.19&0.07&2.65&0.10&2.40&0.05\\
without pl&2.20&0.02&2.61&0.05&2.40&0.02\\
\hline
\end{tabular}
 \caption{\label{Tab: tabela} The table of $\gamma$ power-law exponents for in,
out and
 undirected degree distributions for the eleven largest languages. The exponents
for
 all languages except Polish follow the pattern
 $\gamma_{out}>\gamma_{undirected}>\gamma_{in}$. It is not a surprise that the
Polish
 language exhibits uncommon behavior having in mind its unusual degree
distribution
 depicted in Fig. \ref{Fig:kumulativne}. The average values and corresponding errors of the universal exponents 
are calculated in two ways. The
upper one is calculated as a mean value and a standard deviation of
 different exponents in the sample. The lower
 is calculated with the assumption that all exponents are the same and
differences
 are related to exponent estimation i.e. the error is calculated as the standard error of the mean. It is important to stress that exponents are
not estimated from the degree $k=1$, but from $k_{min}$
 for which the estimated exponent is stable. }
\end{center}
\end{table}
\end{footnotesize}
\normalsize


\begin{figure}[t]
\bigskip
\includegraphics*[width=0.4\textwidth]{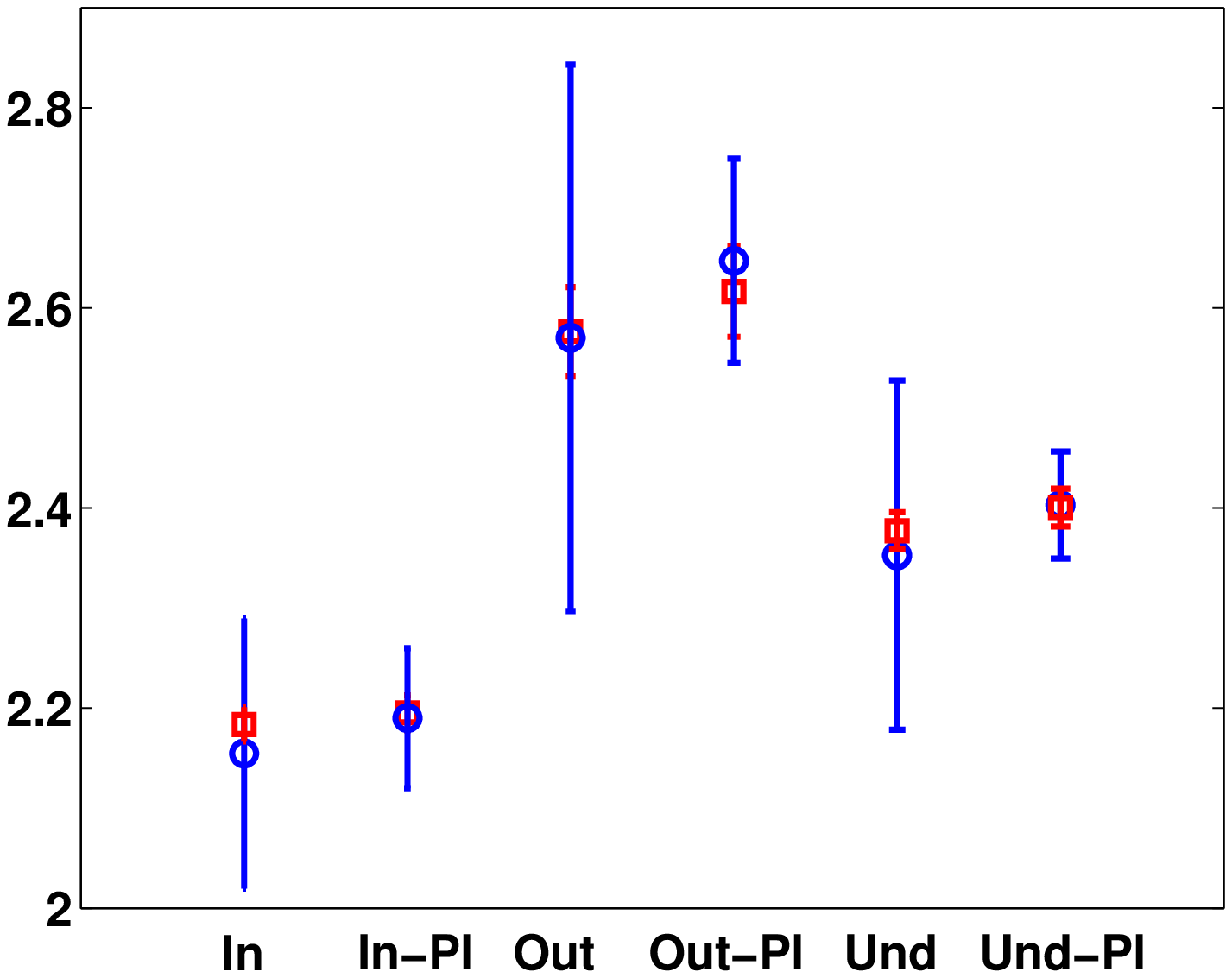}
\caption{\label{Fig:UniExp}(Color online) The obtained universal
exponents for eleven largest languages. The blue (larger bars) represent
mean and the standard deviation of the exponent without the assumption of
universality, while the red (smaller bars) represent the standard
deviation of the exponents with the assumption of universality.}
\end{figure}

\begin{figure}[t]
\bigskip
\includegraphics*[width=0.4\textwidth]{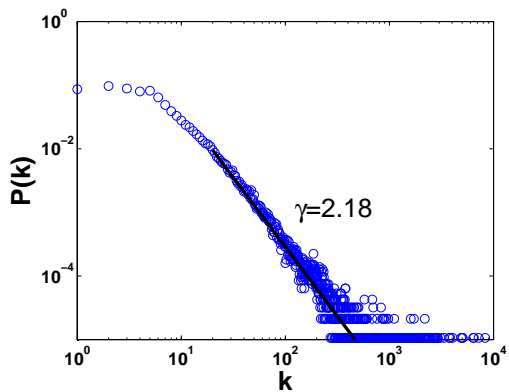}
\caption{\label{Fig:jain}(Color online) The probability distribution
of the in-degree for the Japanese Wikipedia. }
\end{figure}

\begin{figure}[t]
\bigskip
\includegraphics*[width=0.4\textwidth]{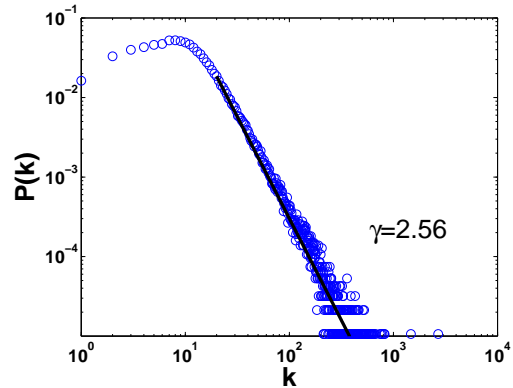}
\caption{\label{Fig:jaout}(Color online) The probability
distribution of the out-degree for the Japanese Wikipedia. }
\end{figure}

One of the most common features of complex networks is the broad degree
probability distribution. The studied Wikipedia networks share this
property with many other complex networks, as clearly shown in Fig.
\ref{Fig:kumulativne}.
The determination of the adequate fitting functional form is a key issue
in the analysis of the broad degree distribution. Many complex networks
have been found to exhibit the scale free nature characterized by the power
law distribution of node degrees $P(k) \sim k^{-\gamma}$. To investigate
a possible power law behavior, we investigated eleven largest languages. The
calculated power law exponents $\gamma$ are presented in the Table \ref{Tab:
tabela}. To estimate the exponents we used the maximum likelihood formula and a nonlinear
fit for the cumulative degree distribution introduced in
\cite{Power}. We did not find any significant size effect on the
exponents $\gamma$. The average $\gamma$ for different languages is
$\gamma_{in}=2.15 \pm 0.13$, $\gamma_{out}=2.57 \pm 0.27$ and
$\gamma_{und}=2.35 \pm 0.17$. Calculated average exponents and their
standard errors were obtained with the assumption that different
realizations of the Wikipedia will have different exponents in the
thermodynamical limit. If their values tended to the same limit,
standard errors would be smaller as depicted in the Fig.
\ref{Fig:UniExp}. While in-degree distributions in general display
the power-law behavior, as an example see Fig. \ref{Fig:jain}, the
power-law nature of the out-degree distribution is much less
expressed (for an example where the power law is clear see Fig.
\ref{Fig:jaout}). Nevertheless, the fat tailed character of the
out-degree distribution is beyond doubt. The estimation for the
out-degree exponent was calculated in a distant tail where the
estimated exponent was sufficiently stable with respect to the
minimal degree of the fitted set $k_{min}$.

In the estimation of average exponents a sample without Polish
language values is also considered, as Polish contains spikes
related to the calendar pages of the Polish Wikipedia. The decision
of the Polish Wikipedia community to heavily interlink calendar
pages using standard templates (e.g. the articles for almost every
year starting with 5 CE link to all days and months of the year and
all years of that century) had enormous repercussions on the degree
distribution of the Polish Wikipedia, as can be clearly observed in
Fig. \ref{Fig:kumulativne}. The exponents for Polish also differ
significantly from other Wikipedia exponents, as can be seen in
Table \ref{Tab: tabela}.

It is interesting to mention that the observed average exponents
agree very well with the WWW exponents for Alta Vista reported in
\cite{N03}.

Alternative distributions we have tested were stretched exponential,
log-normal and the Tsallis distribution. Power law was a significantly
better fit than the other distributions with the exception of the
Tsallis distribution. Because of the larger number of parameters one
needs to estimate for fitting and the unclear phenomenological
origin of the Tsallis distribution we decided to report only the
power law exponents which are commonly understood.

Very recently a paper on the Wikipedia network structure
\cite{wiki}, by Capocci et al., has appeared. The authors use the
complete Wikipedia history to study the growth and structure of
Wikipedia as a complex network. In particular, Capocci et al. find
that the mechanism based on the preferential attachment is adequate
for the description of the Wikipedia growth. The paper also analyzes
Wikipedia topology and assortativity. The comparison of our results
with the results in \cite{wiki} for the node degree probability
distribution exponents shows an agreement for the in-degree
exponents, but reveals a difference in the out-degree exponents
(Capocci et al. report $\gamma_{out}$ between 2 and 2.1 whereas our
estimated average is 2.6).  A possible origin of this discrepancy
could lie in the selected dataset of Wiki pages, or in the power law
exponent estimation techniques. Namely, because the out-degree
distribution is often not a clear power law, one can expect
different results depending on the choice of the minimal degree
$k_{min}$ from which one starts the estimation of the power law
exponent, as well as on the choice of the cut-off degree $k_{max}$
up to which a power law is fitted.

The node degree probability distributions, presented in Fig.
\ref{Fig:kumulativne} and Table \ref{Tab: tabela}, exhibit a high
degree of similarity despite the fact that the corresponding
Wikipedias differ in size by more than an order of magnitude. This
finding supports the assumption that the Wikipedias in different
languages represent realizations of the same process of network
growth. A similar claim is expressed by distinguished members of Wikipedian
communities \cite{WalesMillosh}. The ensemble of all available Wikipedias thus
seems to represent a series of ``snapshots" of the Wikipedia growth process. The
Wikipedias differ significantly in size and degree of development
and, therefore, the ensemble covers many distinct phases of this
growth process.

\subsection{Growth in size}

\begin{figure}[t]
\bigskip
\includegraphics*[width=0.4\textwidth]{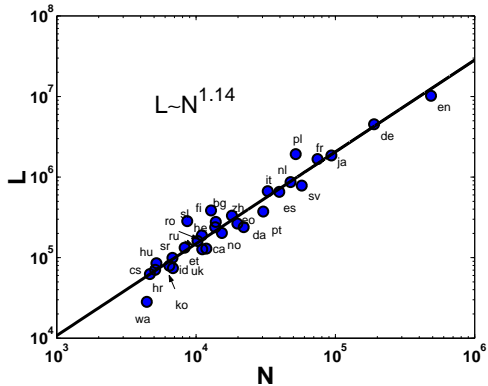}
\caption{\label{Fig: mnogim}(Color online) The number of directed
links plotted against the number of nodes in different Wikipedias.
The growth of $L$ is well described by $N^{1.14}$. This result is
very close to a linear relationship and to determine precisely the
deviation from linearity, should it exist, the study of the history
data for any given language would be necessary.}
\end{figure}

\begin{figure}[h]
\bigskip
\includegraphics*[width=0.4\textwidth]{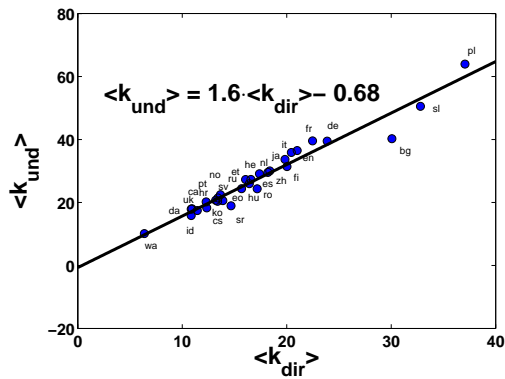}
\caption{\label{Fig:kundkdir}(Color online) The directed and
undirected average degree are in strong correlation across
languages. This implies an important and universal characteristic of
this measure for the Wikipedia network. }
\end{figure}

In light of this,
we report some interesting features of the growth of the number of
crosslinks $L$ with the number of articles $N$ using the said
ensemble of Wikipedias. The growth estimated from different
Wikipedias is $L \sim N^{\alpha}$ with $\alpha =1.14 \pm 0.05$,
which is close to the linear increase of the number of links with
the number of nodes (see Fig. \ref{Fig: mnogim}). A regular
distribution of the points in the plot of Fig. \ref{Fig: mnogim}
further corroborates the hypothesis of a common growth process. A
small difference of the estimated $\alpha$ and 1 is interesting from
the perspective of theoretical models aiming to describe complex
network growth and structure. Namely, a number of models
assume that when a new node is added, approximately the same number
of new links are formed. Such models lead to a linear relationship
between $L$ and $N$ and it is interesting that the ensemble of
Wikipedias is not far from this linear relationship. Clearly, the
models of complex network growth in which the number of links grows
with the number of nodes steeper than linearly are also of interest
from the perspective of explaining Wikipedia network growth and
structure. It would be of special interest to compare the results
obtained from the ensemble of Wikipedias with the ``snapshots" of a
single Wikipedia taken at different stages of its growth. The
estimated growth also implies a slight increase of the average
degree $\langle k_{dir} \rangle \sim N^{\alpha-1}$. The obtained
power law exponents are greater than 2 and therefore we can expect
very limited growth of the average degree, if any.

\subsection{Network topology}

\begin{footnotesize}
\begin{table}[t]
\begin{center}
\begin{tabular}{|l||c|c|c|}
\hline
language&SCC&WCC--SCC&all--WCC\\
\hline\hline
en&85.73&13.17&1.10\\
\hline
de&95.09&4.63&0.28\\
\hline
ja&96.75&2.77&0.49\\
\hline
fr&94.62&5.01&0.37\\
\hline
sv&89.59&9.36&1.04\\
\hline
pl&93.45&6.00&0.55\\
\hline
nl&94.00&5.69&0.31\\
\hline
es&91.55&7.65&0.81\\
\hline
it&86.12&13.60&0.28\\
\hline
pt&87.73&10.83&1.43\\
\hline
zh&89.01&9.22&1.77\\
\hline
\end{tabular}
 \caption{\label{Tab:Topology} The table of network components for 11 largest
languages, in percentages of the total number of nodes.}
\end{center}
\end{table}
\end{footnotesize}
\normalsize

In studying the relative sizes of the regions of the network we used
a more simplified schema than the taxonomy introduced in
\cite{WebTopology} and used in \cite{wiki}. We consider two subsets
of the network: the giant strongly connected component (SCC), where
there is a directed path from every node to another, and the giant
weakly connected component (WCC), where there is an undirected path
between every two nodes. The difference between WCC and SCC includes
the IN, OUT, TENDRILS and TUBES components as well as some nodes
classified by \cite{WebTopology} as disconnected (DISC). The
remaining disconnected nodes are outside the WCC altogether. We
present the relative sizes of these regions in Table
\ref{Tab:Topology}. The sizes of the SCC are on the whole larger
than ones reported in \cite{wiki}. There are two possible ways to
account for this difference. Firstly, our dataset could have been
built using different criteria of selection. Secondly, it dates
after the introduction of categories to Wikipedia. This was a major
structural change, which may have contributed to greater
interconnectivity of all lexicographic topics.

\subsection{Reciprocity}

Another important characteristic of Wikipedia network is the mutual
reciprocity of the links. The average directed degree $\langle
k_{dir} \rangle$ is compared with the average undirected degree
$\langle k_{und} \rangle$ in Fig. \ref{Fig:kundkdir}. There is a
strong correlation between these two moments. Such correlation leads
us to believe that the link reciprocity plays an important role in
the Wikipedia growth process. To understand it better we measured
unbiased mutual reciprocity using the unbiased measure for
reciprocity $\rho$, presented in the paper by Garlaschelli and
Loffredo \cite{Reciprocity}:

\begin{equation}
\rho=\frac{L_{bd}/L-\bar{a}}{1-\bar{a}}.
\end{equation}

Here $L_{bd}$ represents the number of bidirectional links, i.e.
links for which a reciprocal link exists. $L$ is the total number of
directed links and $\bar{a}$ is the density of the links in the
network: $\bar{a}=L/N(N-1)$.
The value of reciprocity for the eleven
largest Wikipedias is $\rho=0.32 \pm 0.05$.

It is interesting to compare the reciprocity of Wikipedia with other
networks that could be very similar to it. The Wikipedias have a
stronger reciprocity than the networks of
associations ($\rho=0.123$ \cite{Reciprocity}) and dictionary terms
($\rho=0.194$ \cite{Reciprocity}),
but smaller than the WWW with $\rho=0.52$ \cite{Reciprocity}.
The difference between the reciprocity of Wikipedia and that of the
WWW will be discussed later in the paragraph on the triad
significance profile. Small Wikipedias show a decrease in
reciprocity with size, which saturates around the reported value,
which is very stable for the largest Wikipedias. This stability of
the measured value suggests that it is a very important quantity for
the description of structure and growth of a Wikipedia-like network.

Reciprocity quantifies mutual ``exchange" between the nodes, and can
be significant in determining whether and to what degree the network
is hierarchical. There have as yet not been many papers dealing with
the origin of reciprocity or network evolution models that capture
this quantity.

\subsection{Clustering}

\begin{figure}[t]
\bigskip
\includegraphics*[width=0.4\textwidth]{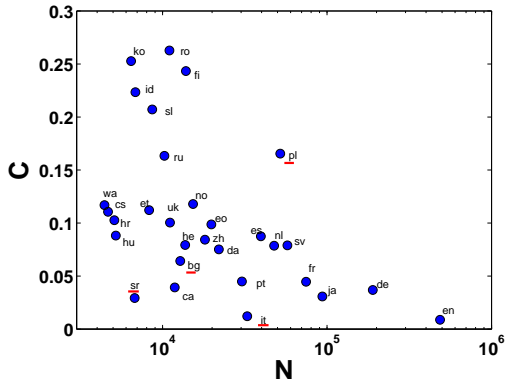}
\caption{\label{Fig:CvsN}(Color online) The dependence of the
clustering coefficient $C$ on the network size $N$. Despite the
significant scattering of the points, it is possible to argue that
Wikipedia clustering coefficient decreases with the network growth.
Wikipedias with unusual degree distributions, underlined in red,
also exhibit a significant deviation from the trend.}
\end{figure}

\begin{figure}[t]
\bigskip
\includegraphics*[width=0.4\textwidth]{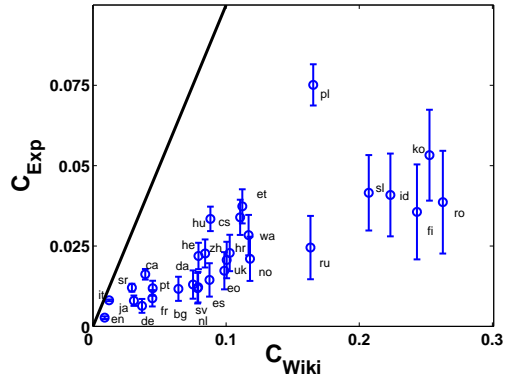}
\caption{\label{Fig:Clustering2} (Color online) Clustering
coefficients of the Wikipedia networks are found to be greater than
one would expect from a random network with the same degree
distribution. From the figure it is obvious that they cannot be
explained as fluctuation from the expected value since the error
bars of expected clustering coefficient, calculated as a standard
deviation of the sample of randomized networks, are far from the
black line which represents $C_{Wiki}=C_{Exp}$.
A great diversity of the measured
clustering coefficients can be explained by the fact that the original
network
is directed, and its undirected representation is missing information important
for the network
growth process.
}
\end{figure}

The clustering coefficient $C$ is one of the most explored values in
complex networks analysis. It is the key quantity in the structure
of undirected networks and represents the local correlation effects
in the node neighborhood.
We calculated the global clustering coefficient, equal to the
probability that the two nodes connected with a path of length 2
also have a mutual link i.e. a path of length 1:

\begin{equation}
C=\frac{3* number\:of\:triangles}{number\:of\:connected\:node\:triplets}.
\end{equation}

In order to determine the clustering coefficient we regarded the
Wikipedia article networks as undirected: every two neighboring
nodes are connected with one undirected link. The relation of the
clustering coefficients to the network size is displayed in Fig.
\ref{Fig:CvsN}. Although the data points are scattered, the general
trend is that the clustering coefficient decreases with the size of
the network. This finding is consistent with other results where
clustering is a finite-size effect \cite{expclust}. It is
interesting to notice that the points which deviate the most from
the general trend, such as Polish or Italian, are also characterized
by deformed degree distributions.

We compared the Wikipedia clustering coefficients to the expected
clustering coefficients of uncorrelated networks calculated from the
known degree probability distribution \cite{expclust}:

\begin{equation}\label{expclust}
C_{exp}=\frac{\left( \left\langle k^2\right\rangle -\left\langle k\right\rangle
\right)^2 }{N\left\langle k\right\rangle ^3}.
\end{equation}

The peculiarities of Polish, Italian, Bulgarian and Serbian degree
distributions have an enormous impact on this calculation. The
expected clustering coefficients obtained by Eq. (\ref{expclust})
for Italian, Bulgarian and Serbian are even greater than 1, which is
clearly impossible. These degree distributions exhibit a peak in the
ultra connected nodes, causing a very large second moment
$\left\langle k^2\right\rangle$, which spoils the results obtained
by analytical reasoning.

An additional contribution to the deviation of Eq. (\ref{expclust})
from the empirical values may lie in the fact that the finite
maximally random networks with a given degree distribution have some
topological constraints (in undirected networks the double links
cannot exist, the nodes cannot link to themselves, the sum of
degrees has to be even). Therefore, these networks are not
necessarily uncorrelated and the underlying assumption of Eq.
(\ref{expclust}) may not be satisfied. It is also plausible that
this effect may be more pronounced in networks with slightly
pathological distributions.

In order to get a better estimate of the expected clustering
coefficient we adapted the algorithm from \cite{Motif} for
randomizing a network with a known degree distribution, and
calculated average clustering coefficients for 100 randomly
generated networks. Comparing this clustering coefficient with the
measured one, we found a significant bias of the Wikipedia networks
to form triangles, see Fig. \ref{Fig:Clustering2}. This is the
result one would expect for a network of definitions, because the
terms referring to one another are likely to refer to further common
terms.

\subsection{Assortativity}

\begin{figure}[t]
\bigskip
\includegraphics*[width=0.4\textwidth]{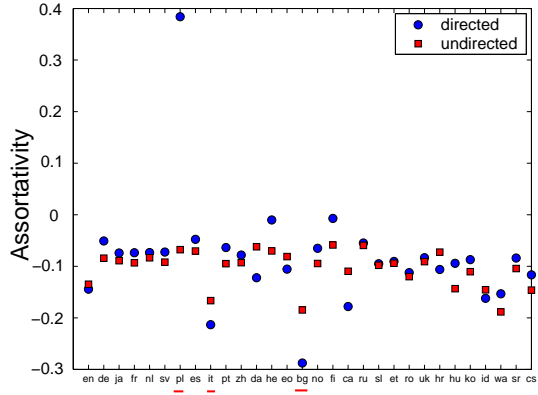}
\caption{\label{Fig: Assort} (Color online) The Wikipedia networks
are found to be slightly disassortative on the whole. The outliers
are marked with red and coincide with the Wikipedias with peculiar
degree distributions.}
\end{figure}

We also calculated the assortativity coefficient of the Wikipedia
network as a global measure of the degree correlations. In
\cite{assort} Newman defines the assortativity coefficient $r$ for
mixing by vertex degree in a directed network as

\begin{equation}
r = {\sum_{jk} jk(e_{jk}-q^{\rm in}_jq^{\rm out}_k)\over
     \sigma_{\rm in}\sigma_{\rm out}}.
\label{rdir}
\end{equation}

Here $e_{jk}$ represents the probability that a randomly chosen
directed link leads out of a node of out-degree $k$ and into a node
of in-degree $j$, $q^{\rm in}_j$ and $q^{\rm out}_k$ are the degree
distributions for in- and outlinks respectively, and $\sigma_{\rm
in}$ and $\sigma_{\rm out}$ are the standard deviations of these
distributions.

This measure describes the likelihood that the nodes of similar
(positive values) or dissimilar (negative values) degrees are
connected, as compared to the random case. The assortativity
coefficient for Wikipedias is slightly negative for all undirected
($r=-0.10 \pm 0.04$) and directed ($r=-0.10\pm 0.05$) Wikipedia
networks except the Polish one, which is strongly assortative in the
case of the directed network ($r=0.38$), as can bee seen on figure
\ref{Fig: Assort}. The small values of the assortativity coefficient
agree well with the more detailed analysis reported by Capocci et
al. in \cite{wiki}. These authors concluded that there was no
significant correlation between the in-degrees of the node. Having
in mind small values of assortativity coefficient we obtained, this
conclusion is very reasonable, but a certain disassortativity is
definitely present in Wikipedia because of the overall negativity of
almost all measured assortativity coefficients.

\subsection{Path lengths}

\begin{footnotesize}
\begin{table}[t]
\begin{center}
\begin{tabular}{|l||c|c|c|}
\hline language&$\langle l_{undir} \rangle$&$\langle l_{dir}
\rangle$&$\langle l_{random} \rangle$\\
\hline\hline
en&3.28&4.90&3.64\\
\hline
de&3.34&4.33&3.30\\
\hline
ja&3.24&4.10&3.26\\
\hline
fr&3.25&4.36&3.04\\
\hline
sv&3.53&4.84&3.52\\
\hline
pl&3.41&4.47&2.61\\
\hline
nl&3.36&4.40&3.18\\
\hline
es&3.38&4.68&3.20\\
\hline
it&3.11&4.77&2.90\\
\hline
pt&3.35&4.65&3.43\\
\hline
zh&3.26&4.36&2.88\\
\hline
\hline
average&3.32&4.53&3.18\\
\hline
error&0.11&0.25&0.30\\
\hline
\end{tabular}
 \caption{\label{Tab: path}
The table of the average path lengths of the undirected paths in WCC
$\langle l_{undir} \rangle$ (arithmetic mean), the average path
lengths of the directed paths in WCC $\langle l_{dir} \rangle$
(harmonic mean) and the expected average path lengths for a random
network (calculated as $\langle l_{random} \rangle=\ln N/\ln\langle
k_{undir} \rangle$), for the eleven largest languages. The displayed
average path lengths exhibit no significant dependence on the size
of the network despite the fact that the studied Wikipedia networks
differ in size by more than an order of magnitude.}
\end{center}
\end{table}
\end{footnotesize}
\normalsize

The path analysis of the Wikipedia networks reveals interesting
results, as shown in Table \ref{Tab: path} for the eleven largest
languages. The studied quantities are the average path length of the
undirected paths in WCC $\langle l_{undir} \rangle$ (calculated as
an arithmetic mean) and the average path length of the directed
paths in WCC $\langle l_{dir} \rangle$ (calculated as a harmonic
mean). For both of these quantities, the largest Wikipedias show no
evidence of scaling of the average path lengths with the network
size.  However, the values of $\langle l_{undir} \rangle$ for all
examined networks are close to the expected average path length for
a random network $\langle l_{random} \rangle=\ln N/\ln\langle
k_{undir} \rangle$, so the Wikipedia networks exhibit small-world
behavior in the original sense.  In addition, the shortest average
path values for the eleven largest languages are very close to one
another, with very small scattering around the average value of the
sample (see Table \ref{Tab: path}). This scattering is considerably
smaller than that of $\langle l_{random} \rangle$.

\subsection{Triad significance profile}

\begin{figure}[t]
\bigskip
\includegraphics*[width=0.4\textwidth]{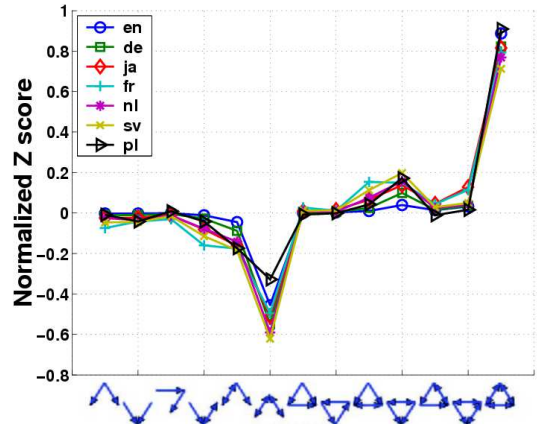}
\caption{\label{Fig:Motifs2}(Color online) The triad significance
profiles of Wikipedias are very similar. The x-axis depicts all possible triads of a directed network, while the 
y-axis represents the normalized Z score for a given triad, given by 
Eq. (\ref{tspeq}).. TSP shapes resemble the
TSP of WWW reported in \cite{Motif}.}
\end{figure}

The last quantity we present in this paper are the triad
significance profiles (TSP), introduced in \cite{Motif}, which
describe the local structure of the networks. Counts of specific
triads (directed three-node subgraphs, shown in Fig.
\ref{Fig:Motifs2} along the abscissa) in the original network are
compared to counts of triads in randomly generated networks with the
same degree distribution.

The significance profile $SP$ is the normalized vector

\begin{equation}\label{tspeq}
{SP}_i = {Z_i \over (\sum_i Z_i^2)^{1/2}}
\end{equation}

of statistical significance scores $Z_i$ for each triad $i$,

\begin{equation}
Z_i = {N_i^{\rm orig} - \langle N_i^{\rm rand} \rangle \over
\sigma_i^{\rm rand}}.
\end{equation}

Here $N_i^{\rm orig}$ is the count of appearances of the triad $i$
in the original network, while $\langle N_i^{\rm rand} \rangle$ and
$\sigma_i^{\rm rand}$ are the average and the standard deviation of
the counts of the triad $i$ over a sample of randomly generated
networks.

In \cite{Motif}, Milo et al. identify superfamilies of networks for
which triad significance profiles closely resemble each other.
Assuming that one can look at the Wikipedia as a representation of
the knowledge network created by many contributors, one could expect
a possible new superfamily of networks. The triad significance
superfamily from \cite{Motif} one would expect to be closest to the
Wikipedia is the one that includes WWW and social contacts.

\begin{figure}[t]
\bigskip
\includegraphics*[width=0.4\textwidth]{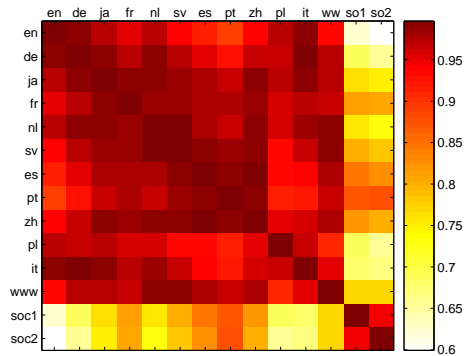}
\caption{\label{Fig:MotifCorr}(Color online) The correlations
between TSPs of the eleven largest languages, the WWW of the nd.edu
domain \cite{BarAlbScience} and the social networks of positive
sentiment between prisoners (soc1) and leadership class students
(soc2) \cite{Motif}.  Wikipedias except for Polish and Italian shown
in order of size.  All Wikipedia profiles and the WWW profile are
pairwise very similar. With the exception of Polish and Italian,
profiles of languages of similar sizes tend to be more closely
correlated. Also, smaller Wikipedias resemble the social networks
better than the larger ones do.}
\end{figure}

The triad significance profile of the largest seven Wikipedias is
depicted in the Fig. \ref{Fig:Motifs2}, and shows common features
found in all examined Wikipedias. These TSPs indeed belong to the
same superfamily as the TSPs of WWW and social contacts reported in
\cite{Motif}, see Fig. \ref{Fig:MotifCorr}. Within this superfamily, the WWW of
nd.edu  exhibits higher correlation with the Wikipedias than the social
networks
do. Since the TSP takes into account the reciprocity of directed links, one
could naively expect that Wikipedia reciprocity would also be very similar to
the WWW's
reciprocity, but we found this is not the case.

\begin{figure}[t]
\bigskip
\includegraphics*[width=0.4\textwidth]{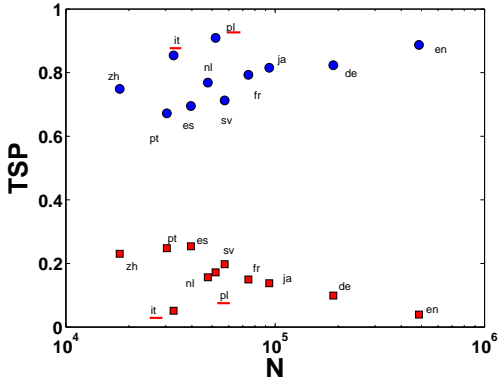}
\caption{\label{Fig:motifsscale}(Color online) The scaling of the
normalized Z score for the most represented triads with the size of
the network. The plot demonstrates that the representation of the
triad 13 (circles) grows, whereas the representation of the triad 10
(squares) falls with the growth of the network. This effectively
means that Wikipedia has a tendency of creating strong
(bidirectional) links for the well connected cliques.}
\end{figure}

The scaling of the triads which are the most represented in the
Wikipedia networks (denoted as 10 and 13) with the network size is
given in Fig. \ref{Fig:motifsscale}. Since both of these triads
represent triangles (see Fig. \ref{Fig:Motifs2}) they contribute to
increasing the clustering coefficient. The Wikipedia TSP thus sheds
additional light on the large clustering of Wikipedia networks, Fig.
\ref{Fig:Clustering2}.

\section{Conclusion}

We have examined the following characteristics of different language
Wikipedia article networks: degree distribution properties, growth,
topology, reciprocity, clustering, assortativity, average shortest
path lengths, and triad significance profiles.  Based on our
results, it is very likely that the growth process of Wikipedias is
universal.  The similarities between Wikipedias in all the measured
characteristics suggest that we have observed the same kind of a
complex network in different stages of development.  We have also
found that certain individual Wikipedias, such as Polish or Italian,
significantly differ from the other members of the observed set.
This difference can be seen most easily in their degree
distributions, but also shows in assortativity, clustering and the
triad significance profile.  In the case of the Polish Wikipedia,
where the discrepancies are the greatest, we have found that they
were caused by an editorial decision involving calendar pages.  This
shows that the common growth process we have observed is very
sensitive to community-driven decisions.

We have shown further that Wikipedia article networks on the whole
resemble the WWW networks.  Specifically, they belong to the TSP
superfamily described in \cite{Motif} that includes WWW and social
networks, and exhibit small-world behavior, with average shortest
path lengths close to those of a random network. In some
characteristics, however, large Wikipedias seem to diverge from the
WWW. Their reciprocity is lower than that of the WWW reported in
\cite{Reciprocity}, and their average shortest path lengths seem to
tend to a stable value.

It is possible that the specific properties of Wikipedias are
related to the underlying structure of knowledge, but also that
their shared features stem from growth dynamics driven by free
contributions, common policies and community decision making.
Whichever the case, the regularities we have found point to the
existence of a unique growth process. These findings in turn support
the method of using statistical ensembles in network research, and,
finally, affirm the role of statistical physics in modeling complex
social interaction systems such as Wikipedia.

{\bf Acknowledgment.} The work of V. Zlati\'c and H.
\v{S}tefan\v{c}i\'{c} was supported by the Ministry of Science,
Education and Sport of the Republic of Croatia. V. Zlati\'c would
like to thank M. Martinis and the members of his project for the
support during the last 4 years. The authors would like to thank D.
Vinkovi\'{c} and P. Lazi\'{c} for important help in computation. We would like to thank R. Milo, N. Kashtan and U. Alon for making the data and the
algorithms from \cite{Motif} available on the Weizmann Institute of
Science web site. We
also thank G. Caldarelli, L. Adamic, P. Stubbs, F.
Mili\v{c}evi\'{c}, and E. Heder for valuable suggestions and
discussions, and K. B\"{o}rner and the Information Visualization
Laboratory, Indiana University, Bloomington, for support and
cooperation. The work of M. Bo\v{z}i\v{c}evi\'{c} was partly
supported by a National Science Foundation grant under IIS-0238261.

\end{document}